\documentclass[iop]{emulateapj}
\usepackage{hyperref}
\usepackage{natbib,graphicx}
\def\scnb{$\rm{S}(3839)$\,}
\def\scnr{$\rm{S}(4142)$\,}
\def\dscnr{$\delta\rm{S}(4142)$\,}
\def\dscnb{$\delta\rm{S}(3839)$\,}

\def\sch{$\rm{CH}(4300)$\,}

\shorttitle{The single-population cluster E 3} 
\shortauthors{Salinas \& Strader}
\journalinfo{To appear in ApJ}

\begin{document}

\title{No evidence for multiple stellar populations in the low-mass Galactic globular cluster E 3\footnotemark[$\dagger$]}
\author{Ricardo Salinas and Jay Strader}
\affil{Department of Physics and Astronomy, Michigan State
  University, East Lansing, MI 48824, USA; rsalinas@pa.msu.edu}

\begin{abstract}
Multiple stellar populations are a widespread phenomenon among
Galactic globular clusters. Even though the origin of the enriched
material from which new generations of stars are produced remains
unclear, it is likely that self-enrichment will be feasible only in
clusters massive enough to retain this enriched material. We searched
for multiple populations in the low mass ($M\sim1.4\times10^4$
M$_{\sun}$) globular cluster E 3, analyzing SOAR/Goodman multi-object
spectroscopy centered on the blue CN absorption features of 23 red
giant branch stars. We find that the CN abundance does not present the
typical bimodal behavior seen in clusters hosting multi stellar
populations, but rather a unimodal distribution that indicates the
presence of a genuine single stellar population, or a level of
enrichment much lower than in clusters that show evidence for two
populations from high-resolution spectroscopy. E 3 would be the first 
bona fide Galactic old globular cluster where no sign of 
self-enrichment is found.
\end{abstract}


\section{Introduction} \label{intro}

 \let\thefootnote\relax
\footnotetext{$^{\mathrm{\dagger}}$ Based on observations obtained at
  the Southern Astrophysical Research (SOAR) telescope, which is a
  joint project of the Minist\'{e}rio da Ci\^{e}ncia, Tecnologia, e
  Inova\c{c}\~{a}o (MCTI) da Rep\'{u}blica Federativa do Brasil, the
  U.S. National Optical Astronomy Observatory (NOAO), the University
  of North Carolina at Chapel Hill (UNC), and Michigan State
  University (MSU).}

Although the fact that globular clusters (GCs) are complex and not
simple stellar populations has been widely acknowledge with the
discovery of multiple stellar sequences in several clusters using
high-precision photometry
\citep[e.g.][]{bedin04,piotto07,milone08,anderson09} and the finding
of a Sodium-Oxygen (NaO) anti-correlation in red giant branch (RGB)
stars from high-resolution spectroscopy
\citep[e.g.][]{drake92,ivans01,gratton01,carretta09a}, star-to-star
variations in the chemical composition of RGB stars and below, hinting
at the existence of these multiple populations, have been known for a
long time thanks to narrow-band imaging and low-resolution
spectroscopy measuring the blue cyanogen (CN) bands
\citep[e.g.][]{osborn71,norris79,norris81,bell83}.

With the exception of a few of the most massive Galactic GCs
\citep{norris96,marino11,yong14}, GCs are homogeneous in iron-peak
elements to within $\sim$0.1 dex \citep{carretta09b,willman12},
meaning that the pollution to the ISM necessary to produce distinct
populations cannot come from supernovae. Instead, a number of other
mechanisms have been discussed involving stellar winds from rotating
low-metallicity stars \citep{maeder06}, winds coming from Wolf-Rayet
stars \citep{smith06}, massive interacting binaries \citep{demink09}
 polluting circumstellar discs of pre-main sequence stars
  \citep{bastian13}, or novae \citep{maccarone12}, although the most
heavily discussed mechanism is pollution by the ejecta of
intermediate-mass asymptotic giant branch (AGB) stars
\citep[e.g.][]{cottrell81,ventura01}. In the case of clusters in the
Magellanic Clouds, the presence of extended MS turn-offs (MSTO) in
intermediate-age clusters has been interpreted as the presence of
multiple stellar populations \citep[e.g.][]{mackey08}, although this
intepretation has been highly disputed
\citep[e.g.][]{bastian15,brandt15} Extended MSTOs are visible in
clusters with masses down to $\sim 10^4 M_{\sun}$
\citep{milone09,goudfrooij11}. Clusters below this limit would host
single stellar populations \citep{conroy11}.

Regardless of the origin of the enriched material, a key aspect is the
ability to retain it in order to form new generations of stars. While
the most massive clusters could retain SNe ejecta (e.g. $\omega$Cen),
less massive GCs would retain only the more gentle outflows produced
by the mechanisms mentioned above. This leads to the natural question
of whether there is a mass limit below which no ejected material could
be retained and hence genuine single-population GCs be produced
\citep{caloi11}.

\subsection{The low mass cluster E 3}

E 3 \citep[RA=09:20:57.07, Dec=--77:16:54.8][]{goldsbury10} is one of
the sparsest ($r_h=4.94$ pc) and faintest GCs in our Galaxy
\citep{lauberts76}. A metal-rich cluster
\citep[{[}Fe/H{]}=--0.74][]{carretta09a}, is considered as part of the
old GCs in our Galaxy \citep{marin09}, although its age could be as
low as 2 Gyr less than 47 Tuc \citep{sarajedini07}, where the
difficulties in establishing a precise age mostly stem from the
uncertainties in its distance due to the lack of horizontal branch
stars: while older measurements indicate a distance with
$(m-M)_0=13.19$ \citep{harris96}, \citet{sarajedini07} measures a
larger distance modulus of 14.54, value we adopt throughout this
paper.

E 3 is also known for its very prominent binary star main sequence
\citep{mcclure85,veronesi96}. \citet{milone12} studied 59 Galactic GCs
finding E 3 had the highest binary fraction in the sample.

Most relevant for this paper, E3 is among the lowest mass GCs in
our Galaxy. With an apparent magnitude of $g=10.79$
\citep{vanderbeke14} and a $M/L_g=2.83$ (for a 10 Gyr single
  stellar population using the \citealt{maraston05} models), E3 has a mass of
$\sim1.4\times10^4$M$_{\sun}$. Therefore, given its relatively short
distance and low mass, E 3 provides one of the best targets to probe
the existence of a mass limit for self-enrichment in Galactic GCs.

Finally, during the referee process of this paper,
\citet{delafuente15} presented high-resolution spectroscopy of 9 stars
in the E 3 field, judging only two as cluster members based on the
derived temperatures. From these two members, they find a radial
velocity for the cluster $v_r=45\pm5$ km s$^{-1}$, and based on thr
proper motion of these two stars also derive a very high tangential
velocity of $382\pm79$ km s$^{-1}$. We discuss their radial velocity
compared to our own measurement.

                                                                     
\section{Observations and data reduction} \label{obs}

\subsection{SOAR/Goodman imaging}

Optical imaging of E 3 was conducted with the Southern Astrophysical
Research (SOAR) 4.1m telescope located in Cerro Pach\'on, Chile. The
imaging mode of the Goodman spectrograph \citep{clemens04} was used to
obtain short exposures with each Kron-Cousins $B$ and $R$
filters on the night of October 1, 2013. Further $BV$ imaging was
obtained on the night of January 14, 2015. Goodman provides a circular
field of view of $7.2\arcmin$ in diameter in imaging mode.

Overscan subtraction, flat fielding and image alignment were applied
with standard tools within \textsc{iraf}. FWHM measured on the
combined images was $\sim 2.2\arcsec$ for the December 2013 images and
$\sim1\arcsec$ for the January 2015 run.

\begin{figure}
\includegraphics[width=0.48\textwidth]{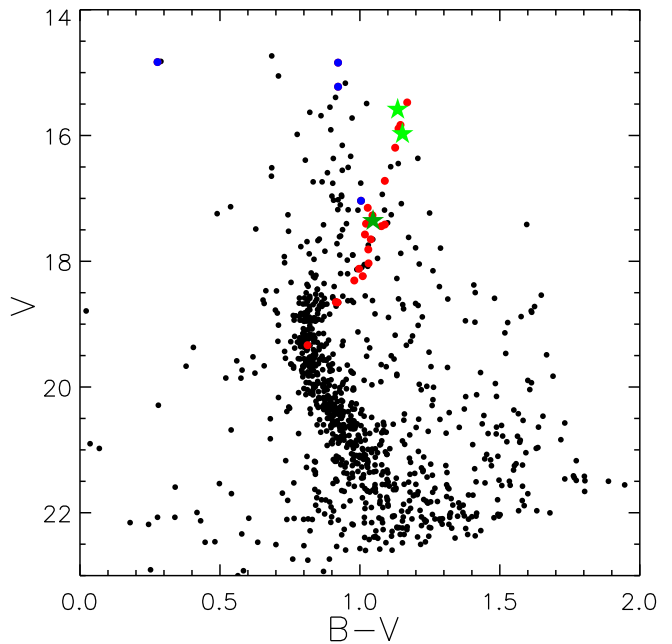}
\caption{$BV$ CMD of E 3. Member stars are indicated in red circles,
  while blue circles indicate stars judged as non-members based on
  their position in the CMD. The dark green star indicates the member
  star observed in both masks, while the light green star near the tip
  of the RGB were observed in a higher-resolution mode
  (Sect. \ref{sec:velocity})}
\label{fig:cmd}
\end{figure}

A catalog of sources produced with SExtractor \citep{bertin96} was fed
into \textsc{daophot} \citep{stetson87} for measuring aperture
photometry. Given the sparsity of the field, we found 5-pixel aperture
photometry sufficient for our goals. Aperture corrections were
established using 7 isolated secondary Stetson standards$^1$
\footnote{$^1$www3.cadc-ccda.hia-iha.nrc-cnrc.gc.ca/community/STETSON/}
which were also used to put magnitudes into the standard system. A
color-magnitude diagram based on the January 2015 observations is
shown in Fig. \ref{fig:cmd}.

\subsection{SOAR/Goodman multi-object spectroscopy}

Multi-object spectroscopy for selected sources was carried out during
the commissioning run of the multi-object (MOS) capability of the
Goodman spectrograph on the SOAR Telescope, during the night of
December 18, 2013 (``mask 1''). Further observations were taken on
February 17, 2015 (``mask 2''). Masks were prepared using the Slit
Designer software developed at the University of North Carolina. An
astrometric solution for the Goodman pre-images was found using the
web-based tool
astrometry.net$^2$ \footnote{$^2$http://nova.astrometry.net/}
\citep{lang10}.

The Goodman MOS provides a fov of $3\arcmin\times5\arcmin$. Mask 1
consisted of 13 slits, but only 12 stars could be extracted. Mask 2
had 14 slits covering 16 stars. Exposure times were $8\times900$ s and
$4\times900$ s for masks 1 and 2, respectively. The 930 l$/$mm grating
centered at 4100\AA\, was used for both masks, ensuring the presence
of the CN 3883 \AA\, and 4215 \AA\, bands, CH 4300 \AA\, and Ca II H
\& K features in all spectra. H$\beta$ was also present in the
wavelength range in most of the cases. Iron Argon comparison lamps
were taken about every hour of observation. This setup gives a
  resolution of $\sim$ 2.7 \AA\, FWHM.

Standard reduction procedures were conducted using \textsc{iraf}. 
Wavelength calibration achieved an rms of $\sim 0.03$\AA. Individual 
spectra were visually checked for defects before average. In the case 
of cosmic rays or other blemishes in the target wavelength ranges (see 
Sect. \ref{sec:index}), the individual spectrum was removed before 
combination.

\subsection{Radial velocities and SOAR longslit spectroscopy} \label{sec:velocity}

As our observations were taken during commissioning of the Goodman
multi-slit mode, the mask alignment was imperfect, and due to the blue
wavelength range, no telluric absorption lines were present in the
spectra to correct for the effects of slit miscentering. Therefore
these spectra were unsuitable for the measurement of precise radial
velocities.

To nail down the systemic velocity of the cluster, on 2015 June 13 we
obtained a 600 sec longslit spectrum of two bright giants whose
position in the color-magnitude diagram (light green stars in
Fig. \ref{fig:cmd}) suggested they were very likely to be cluster
members. We used SOAR/Goodman with a 2400 l mm$^{-1}$ grating and a
1.03\arcsec slit, covering a wavelength range of $\sim 5100$--5600
\AA\ at a resolution of about $0.75$ \AA. The spectra were reduced in
the standard manner and radial velocities derived through
cross-correlation with spectra of bright stars of similar spectra type
taken with the same setup. 

The radial velocities of the two stars are consistent to within 4 km
s$^{-1}$ and the mean heliocentric radial velocity is 8.9 $\pm$ 2.8 km
s$^{−1}$. This velocity is in stark contrast with the velocity derived
by \citet{delafuente15}, $v_r = 45 \pm 5$ km s$^{-1}$, which appeared
during the referring process of this paper. Even though the cluster is
known for undergoing tidal stripping \citep{vdb80} and having a large
binary fraction \citep{milone12b}, both factors that would increase the
velocity dispersion of the cluster, the very marked difference between
the two measurements suggests that one is not correct. This may be due
to the presence of foreground stars in the sample. In the absence of
additional information, we consider the systemic velocity of E3 to be
uncertain. Fortunately, this issue can be easily settled with future
high-resolution spectroscopy.

\section{Index definition and measurements}\label{sec:index}

\begin{figure}
\includegraphics[width=0.48\textwidth]{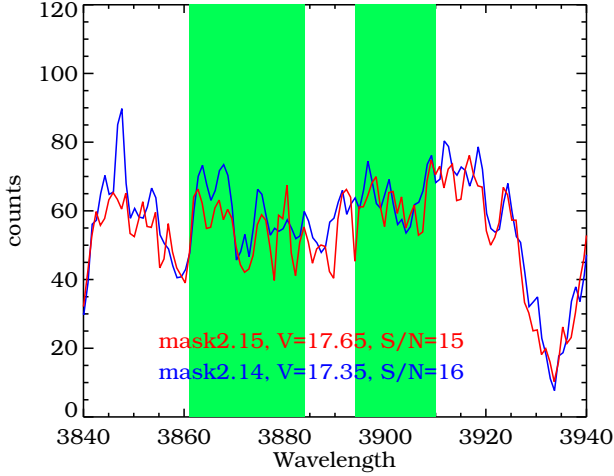}
\caption{The definition of the S(3839) index. The shaded region on the
  left shows the CN region, while the right shaded region is the
  adjacent continuum. The spectra of two stars with similar S/N is
  also shown.}
\label{fig:index_s3839}
\end{figure}

\begin{figure}
\includegraphics[width=0.48\textwidth]{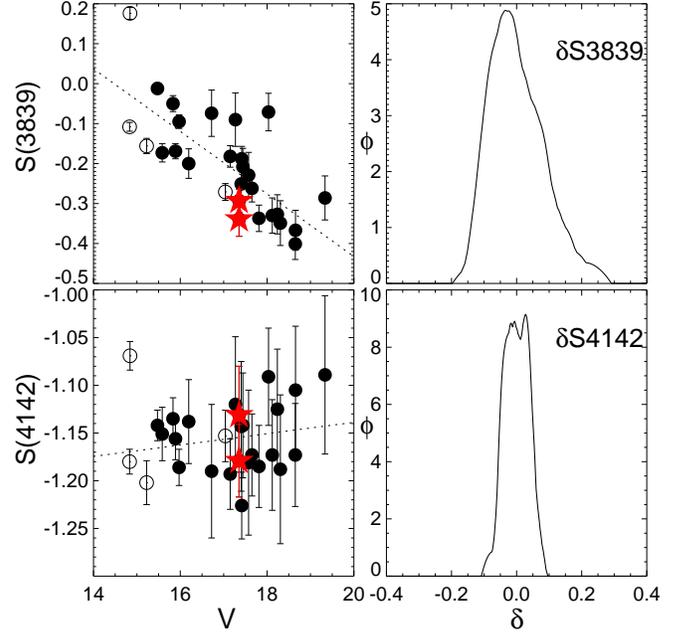}
\caption{\scnb and \scnr indices of RGB stars in E 3. Left panels show
  the uncorrected indices. Open symbols indicate non-member stars
  based on their CMD position. Red stars indicate duplicated
  measurements. Dashed lines indicate a robust linear fit to each
  distribution. Right panels indicate the density distribution of the
  corrected indices \dscnb and \dscnr.}
\label{fig:indices}
\end{figure}

CN and CH abundances (in rigor, line \textit{strenghts}, although we
will use the word abundance hereafter) were measured through spectral
indices, which compare the flux value inside a window bracketing a
spectral feature, with one or two adjacent windows which define a
pseudo-continuum. We adopted the index definitions of
\citet{harbeck03} in order to compare with other results from the
literature (Sect.  \ref{sec:results}),

\begin{eqnarray}
\rm{S}(3839) & = & -2.5 \log{\frac{f_{3861-3884}}{f_{3894-3910}}}, \\
\rm{S}(4142) & = & -2.5 \log{\frac{f_{4120-4216}}{0.4\,f_{4055-4080} + 0.6\,f_{4240-4280}}}, \\
\rm{CH}(4300)& = & -2.5 \log{\frac{f_{4285-4315}}{0.5\,f_{4240-4280} 
+ 0.5\,f_{4390-4460}}},
\end{eqnarray}
where each term is the sum of the flux in counts within the specified
wavelength range. The uncertainties were measured assuming Poissonian
noise for each flux measurement. Fig. \ref{fig:index_s3839} shows
  the definition of the S(3839) index over two sample spectra.

\begin{figure}
\includegraphics[width=0.48\textwidth]{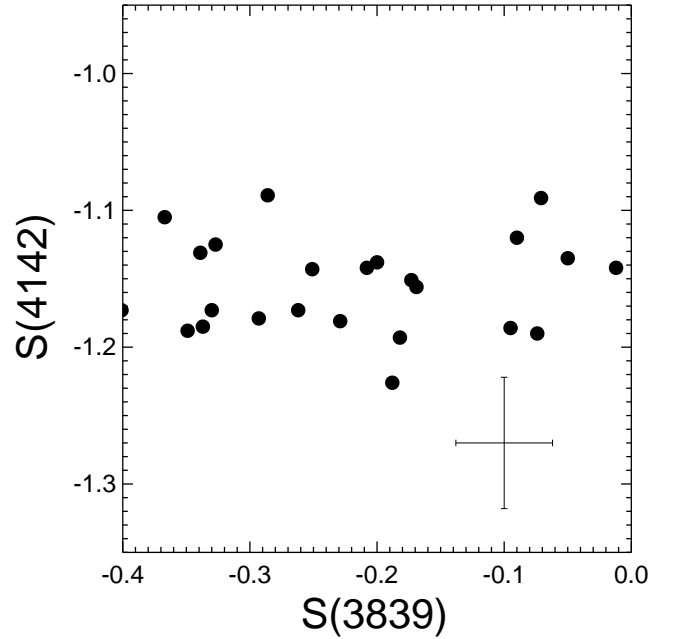}
\caption{Comparison of \scnb and \scnr indices. Median errors are
  depicted as bars in the lower right corner.  While the \scnb
  distribution is significantly broader than its median error, the
  \scnr is consistent with its internal errors.}
\label{fig:comparison}
\end{figure}

Table \ref{tab:stars} shows the measured values and uncertainties for
the indices \scnb, \scnr and CH(4300). Star 3 in the first mask was
observed again as star 11 in the second mask (indicated with a star
symbol in Figs. \ref{fig:cmd} and \ref{fig:indices}); the difference
between the index measurements for this star is of the order of the
calculated Poissonian errors, providing evidence for the consistency
of our measurements between masks.

The intensities of the CN and CH indices are not only a function of
chemical abundance, but also of temperature and gravity. To first
order, these dependences can be removed using the proxies of color
\citep{harbeck03} or luminosity \citep[e.g.][]{norris81,kayser08}, by
fitting a curve to the lower envelope of the distribution
\citep[e.g.][]{harbeck03} or by finding a ridge line that describes
well the data \citep[e.g.][]{pancino10}. In this case we adopt an
approach similar to \citet{pancino10}, by robust fitting of a straight
line to the indices as function of $V$ magnitude (see
Fig. \ref{fig:indices}). The right panels of Fig.~\ref{fig:indices}
show the density distribution of the corrected indices, \dscnb and
\dscnr, defined as the distance to the fitted line.

Even though the \scnr feature is significantly weaker than the \scnb
\citep[e.g.][]{norris81}, it benefits from having a $\sim$50\% higher
S/N (Table \ref{tab:stars}). The sensitivity and reliability of the
indices can be tested by comparing the widths of their distributions
to the measured errors. Fig. \ref{fig:comparison} shows the \scnr
index as function of \scnb. While the spread in \scnr values is
consistent with the uncertainties, the spread in \scnb is larger than
the uncertainties, indicating a bigger sensitivity. Even though the
\scnr index shows a hint of bimodality (lower panels in
Fig. \ref{fig:indices}), there are two reasons to discount it as not 
significant: first, the spread in the \scnr index is consistent with 
that expected on the basis of the measurement uncertainties; second, 
the single star with multiple measurements ``switches'' between the 
two populations. Thus, while bimodality in \scnr might well be 
present, we cannot claim its presence on the basis of these 
observations. Rather, we will focus the analysis on the stronger 
\scnb index, as many studies before 
\citep[e.g.][]{harbeck03,lardo12,lim15}.

\section{Results}\label{sec:results}
\begin{figure}
\includegraphics[width=0.48\textwidth]{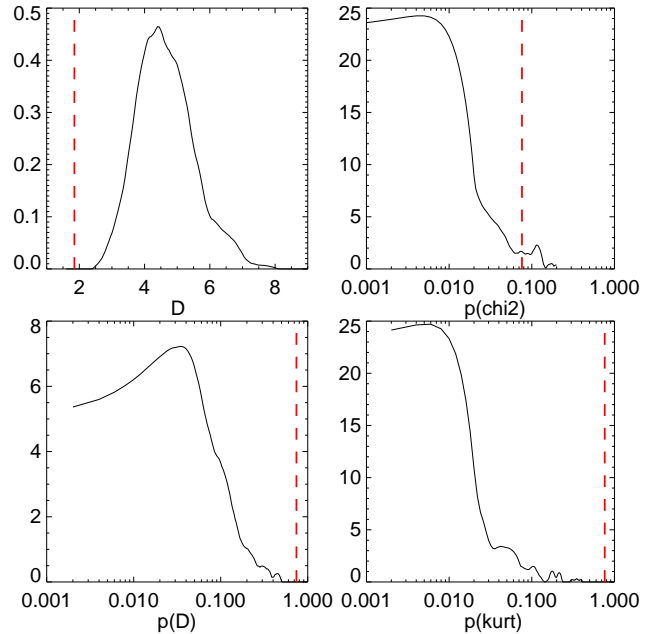}
\caption{GMM statistics for a ``blurred'' bimodal distribution where
  the peaks in the \scnb distribution are separated by 0.25 mag (see
  text for details). For each statistic the vertical red dashed line
  indicates the GMM result for the original sample.}
\label{fig:gmm}
\end{figure}

\subsection{A unimodal CN abundance?}

Fig. \ref{fig:indices} (top right panel) shows the density
distribution of the \dscnb measurements, obtained using a kernel
density estimator with an Epanechnikov kernel. The density
distribution does not show obvious evidence for bimodality, but rather
shows a behavior close to Gaussianity. To further test this visual
impression, we use a Gaussian Mixture Modeling (GMM) as implemented by
\citet{muratov10}. GMM makes a maximum likelihood estimation of the
parameters associated with the selected number of Gaussians (two in
this case), calculating uncertainties in these parameters via
non-parametric bootstrap. It further calculates the distance between
the peaks,
$D=|\mu_{1}-\mu_{2}|/[(\sigma_{1}^2+\sigma_{2}^2)/2]^{1/2}$, and the
kurtosis of the sample. Finally, using a parametric bootstrap it
calculates the probability that the observed distribution is drawn
from a single Gaussian. 

All the quantities measured by GMM are inconsistent with a bimodal
distribution. GMM finds $D=1.85$, below $D=2$ which is considered as a
clear separation between the peaks \citep{ashman94} and a slightly
negative kurtosis of $-0.102$, both quantities with probabilities
0.846 and 0.763 of being obtained from a unimodal distribution,
respectively.

Is it possible that a true underlying bimodal distribution in \scnb is
hidden by observational errors and the relatively low S/N? We tested
this hypothesis with two methods, generating a mock bimodal
distribution, blurring it with the observational errors, and a second
approach measuring the \scnb index in SDSS spectra of the GC M71 with 
artificially reduced quality.

In the first method, we generated two random Gaussian distributions
separated by a conservative $\Delta$ \dscnb=0.25 mag (see below) and
with $\sigma=0.06$ mag each, that is, slightly larger than the
measured median uncertainty in the \scnb values. The value of the
separation between the input Gaussians is also smaller than the usual
separation between CN-rich and CN-weak stars, which is close to 0.4
mag \citep[e.g.][]{norris81,kayser08,lim15}. Each sample consisted of
23 objects randomly placed on either Gaussian. 200 samples were
generated and run through GMM.
\begin{figure}
\includegraphics[width=0.48\textwidth]{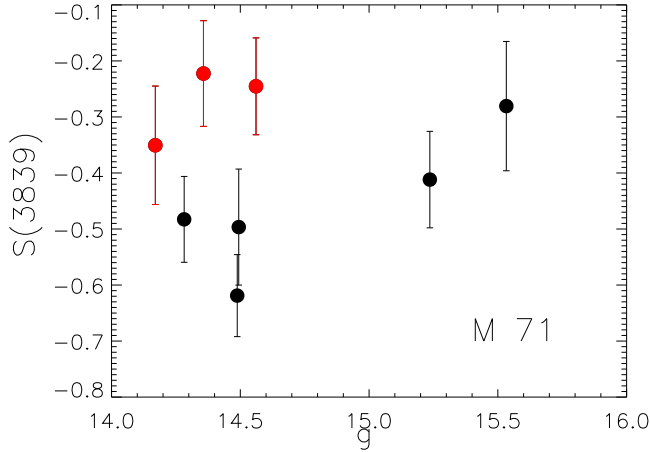}
\caption{CN abundance in the GC M 71. Filled symbols indicate the
  measured \scnb on SDSS spectra (red circles indicating CN-rich
  stars), while error bars indicate the full range of values obtained
  from the same spectra when downgraded to S/N=10.}
\label{fig:m71}
\end{figure}

Fig. \ref{fig:gmm} shows the results of this exercise, where the
vertical dashed lines represent the values obtained when GMM was
applied to the original sample. The top left panel gives the
distribution of distances between the peaks for the 200 generated
bimodal samples. The peak separation is always higher than the
observed value of 1.85. The three other panels show the probability of
obtaining the measured $\chi^2$, peak distance and kurtosis from a
unimodal distribution. These distributions are significantly different
from the ones measured in the original sample, strongly rejecting
bimodality with peak separations of 0.25 mag and above, and supporting
a genuine unimodal distribution of the measured \scnb values.

In the second method, we wanted to test directly the influence of low
S/N in the reliability of the \scnb index measurement. To this end, we
retrieved SDSS/SEGUE \citep{yanny09} spectra of 9 RGB stars belonging
to the GC M71, which has a very similar metallicity to E 3,
[Fe/H]$=-0.82$ \citep{carretta09a}. These data were already used by
\citet{smolinski11}, finding bimodality in the \scnb index, using the
\citet{norris81} definition, slightly different than the
\citet{harbeck03} definition used throughout this paper. We downgraded
the quality of the spectra to S/N=10 and measured the \scnb index in
100 realizations of the original spectra with added
noise. Fig. \ref{fig:m71} shows the results of this approach. Solid
circles show the measurements of the original spectra, while the error
bars indicate the full range of measured values from the Montecarlo
procedure. Despite the low number of stars and some confusion for the
brightest stars, bimodality would remain clear under low S/N
conditions.

Both methods show that a bimodal distribution with the separation
expected for a metal-rich cluster will not become unimodal under
relatively low S/N conditions, supporting the idea that the RGB of E 3
comprises a single stellar population.

\section{Discussion.}

Even though the CN bimodality has been found in a large number of
clusters \citep[e.g.][]{alves08}, clusters with no sign of a CN-rich
population are not without precedent. \citet{kayser08} studied CN
abundances in 8 GCs, finding that two of them, Terzan 7 and Palomar
12, do not present CN-rich stars. These two clusters are also
anomalous in the sense that the widespread NaO anticorrelation is
neither present \citep{sbordone07,cohen04}, although this result
  is based only on a handful of stars. Perhaps most importantly, both
clusters are associated to the Sagittarius dwarf
\citep[e.g.][]{dacosta95,dinescu00}. This led \citet{kayser08} to
suggest that the environment in which the clusters are formed would
influence the presence of CN variations.

Another case is given by the low-metallicity cluster NGC 6397
([Fe/H]$=-1.99$) which also serves as a cautionary tale: while no CN
bimodality was found using low resolution spectra and narrow-band
imaging \citep{lim15}, and also no multiple stellar populations were
found using high quality multi-color ground-based photometry
\citep{nardiello15}; high-precision HST photometry revealed the
presence of two main sequences \citep{milone12b}. This apparent
contradiction can be explained by the weaker CN absorption expected in
low-metallicity clusters \citep[e.g.][]{smolinski11}, which is not the
case for E 3.NGC 6397 has also a well-studied Na-O
  anti-correlation \citep[e.g.][]{gratton01,carretta09a}.

Based on detailed chemical abundances from high-resolution
spectroscopy of 9 RGB stars, \citet{villanova13} claimed Ruprecht 106 
as the first single-population cluster in our Galaxy. Even though its 
luminosity implies a larger present-day mass
than E 3, Rup 106 is also regarded as a GC of extragalactic origin
\citep{lin92,pritzl05,villanova13}. If environment plays a role in the
generation of multiple stellar populations \citep{kayser08}, it is not
unlikely the mass limit for self-enrichment will also be a function of
environment.
\begin{figure}
\includegraphics[width=0.48\textwidth]{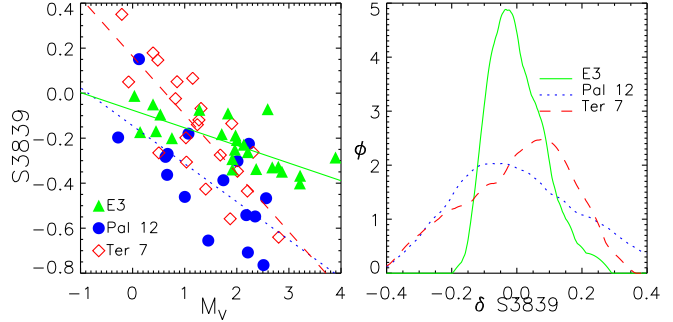}
\caption{A comparison between the \dscnb distribution of E 3 and two
  clusters of similar metallicity, Terzan 7 and Palomar 12. The spread
  in the E 3 values is even narrower than in these two clusters
  considered as unimodal \citep{kayser08}. }
\label{fig:comparison2}
\end{figure}

\subsection{A mass limit for self-enrichment?}

Regardless of which mechanism expells processed material into the ISM 
in GCs, the potential of the cluster (and thus its mass and size)
will determine how much material can be retained. Therefore it is 
likely that little or no material would be retained for 
self-enrichment below some cluster mass. The absence of abundance 
variations in the less massive open clusters 
\citep[e.g.][]{norris85,martell09,carrera13,bragaglia14}, is 
consistent with the existence of such a mass limit, though there may 
be other relevant factors, such as the physical conditions in the 
immediate environment of the forming cluster.

On the basis of the detailed abundances of a sample of 21 GCs and a
comparison to open clusters, \citet{carretta10} suggest a limit of
$M_V = -5.1$ (about $4 \times 10^4 M_{\sun}$ for an old stellar
population), above which all GCs appear to show evidence of
self-enrichment via the Na--O anti-correlation. This limit was chosen
to separate the low mass GC Palomar 5 (which does show multiple
populations) from the open clusters and the GCs Terzan 7 and Palomar
12, which do not show multiple populations. As mentioned before,
besides their low masses, Terzan 7 and Palomar 12 also share an
unusual characteristic, which is that they are much younger than the
bulk of the Milky Way GC system and are inferred to have been accreted
as part of the Sgr dwarf galaxy. Thus, it is unclear whether these GCs
did not self-enrich because of their low masses and large sizes, or
due to some other factor. We also note that E 3 is widely considered
as a genuine Galactic cluster, not associated to any dwarf galaxy or
stream \citep{forbes10}. \citet{carretta10} consider it as a part of
the disk/bulge subsystem of GCs based on its position inside the
Galaxy.

In our view it is premature to make sweeping statements about the 
existence of a mass limit for self-enrichment. First, little data 
exists for low-mass GCs at all: there is a clear observational need to 
both improve data on E 3 (including high-resolution abundance 
measurements for bright giants) and to obtain low-resolution 
spectroscopy or medium-band photometry on other low-mass Milky Way 
GCs. One happy consequence of improved searches for satellites of the 
Milky Way has been the discovery of likely new low-mass GCs. The other 
problem is in interpretation. It is challenging to relate the 
present-day mass of an old, low-mass GC to its initial mass, though 
with improved Galactic models and cluster orbits from Gaia, the 
modeling of the mass loss from individual Milky Way GCs should be 
improved. A similar challenge exists for the initial structural 
parameters of the cluster, which can affect the central escape 
velocity at early times.

\section{Summary and conclusions}

We have presented SOAR/Goodman MOS spectroscopy of 23 RGB stars in the
metal-rich, low-mass GC E3. We measured the blue CN absorption at 3883
\AA finding no evidence of an intrinsic spread in CN line strength. E3
is the first old Galactic GC consistent with a single stellar
population.

\acknowledgments 

We thank an anonymous referee for comments that improved the paper. We
thank Graeme Smith for useful conversations and Nathan Bastian,
Timothy Brandt, Iskren Georgiev and Ricardo Schiavon for comments on a
previous version of this manuscript. We also thank Christopher Clemens
and Josh Fuchs for their assistance during the first Goodman/MOS
run. We acknowledge NSF support through grant AST-1514763.

Funding for the SDSS and SDSS-II has been provided by
the Alfred P. Sloan Foundation, the Participating Institutions, the
National Science Foundation, the U.S. Department of Energy, the
National Aeronautics and Space Administration, the Japanese
Monbukagakusho, the Max Planck Society, and the Higher Education
Funding Council for England. The SDSS Web Site is
http://www.sdss.org/.

The SDSS is managed by the Astrophysical Research Consortium for the
Participating Institutions. The Participating Institutions are the
American Museum of Natural History, Astrophysical Institute Potsdam,
University of Basel, University of Cambridge, Case Western Reserve
University, University of Chicago, Drexel University, Fermilab, the
Institute for Advanced Study, the Japan Participation Group, Johns
Hopkins University, the Joint Institute for Nuclear Astrophysics, the
Kavli Institute for Particle Astrophysics and Cosmology, the Korean
Scientist Group, the Chinese Academy of Sciences (LAMOST), Los Alamos
National Laboratory, the Max-Planck-Institute for Astronomy (MPIA),
the Max-Planck-Institute for Astrophysics (MPA), New Mexico State
University, Ohio State University, University of Pittsburgh,
University of Portsmouth, Princeton University, the United States
Naval Observatory, and the University of Washington.

{\it Facility:} \facility{SOAR}

\bibliography{e3}

\clearpage
\begin{deluxetable}{lcccccrrcrrcl}
 \tabletypesize{\scriptsize}
 \tablecaption{Data for program stars \label{tab:stars}}
 \tablehead{
  \colhead{ID} & \colhead{$\alpha_{2000}$} & \colhead{$\delta_{2000}$} 
& \colhead{$V$} & \colhead{$(B-V)$} & 
\colhead{S(3839)} & \colhead{$\delta$\scnb} 
&\colhead{S/N}\tablenotemark{a} & \colhead{S(4142)} 
& \colhead{$\delta$\scnr} &\colhead{S/N}\tablenotemark{b}&
  \colhead{\sch} &  \colhead{Member?}\\
  & & & mag&mag& }

\startdata
m1.1 & 140.1972 & -77.3133 & 15.224 & 0.922 &--0.156$\pm$0.019 &-0.097 & 24.6 &--1.202$\pm$0.023&-0.035& 38.6 &1.009$\pm$0.019&N\\
m1.2 & 140.1748 & -77.3129 & 14.832 & 0.277 &--0.108$\pm$0.011 &-0.080 & 64.5 &--1.180$\pm$0.013&-0.010& 90.6 &1.005$\pm$0.011&N\\
m1.3\tablenotemark{c} & 140.2880 & -77.2892 & 17.354 & 1.047 &--0.293$\pm$0.029 &-0.063  & 13.8 &--1.179$\pm$0.038&-0.024& 22.4 &1.056$\pm$0.032&Y \\
m1.4 & 140.2618 & -77.2917 & 18.237 & 1.010 &--0.327$\pm$0.049 &-0.032  & 3.2  &--1.125$\pm$0.063&0.024& 7.5  &1.048$\pm$0.053&Y \\
m1.5 & 140.2160 & -77.2920 & 17.442 & 1.077 &--0.208$\pm$0.046 &0.024 & 8.1  &--1.142$\pm$0.055&0.012& 10.6 &1.148$\pm$0.046&Y\\
m1.6 & 140.2543 & -77.2762 & 17.150 & 1.028 &--0.182$\pm$0.027 &0.027 & 13.1 &--1.193$\pm$0.037&-0.037& 17.8 &0.987$\pm$0.032&Y\\
m1.7 & 140.2197 & -77.2771 & 18.120 & 0.997 &--0.330$\pm$0.044 &-0.045  & 4.2  &--1.173$\pm$0.058&-0.023& 7.9 &1.024$\pm$0.048&Y\\
m1.8 & 140.1445 & -77.2842 & 15.584 & 1.135 & --0.173$\pm$0.023&-0.086 & 23.2 &--1.151$\pm$0.028&0.014& 35.5 & 1.072$\pm$0.023&Y\\
m1.9 & 140.1167 & -77.2829 & 17.403 & 1.022 &--0.251$\pm$0.058 &-0.022 & 5.3  &--1.143$\pm$0.068&0.011& 7.4& 1.048$\pm$0.056&Y\\
m1.10 & 140.1340 & -77.2736 & 14.841 & 0.922 & +0.176$\pm$0.015&0.205 &50.7   &--1.069$\pm$0.015&0.100& 80.6& 1.082$\pm$0.012&N\\
m1.11 & 140.1085 & -77.2686 & 15.832 & 1.145 &--0.050$\pm$0.020 & 0.056  &18.8  &--1.135$\pm$0.022&0.029& 43.6&  1.117$\pm$0.019&Y\\
m1.12 & 140.0702 & -77.2658 & 15.889 & 1.138 & --0.169$\pm$0.019 &-0.058 &24.8 &--1.156$\pm$0.022&0.007& 40.5& 1.138$\pm$0.018&Y\\

m2.1 & 140.2114 & -77.3321 & 17.270 & 1.045 & --0.090$\pm$0.067 & 0.129 & 3.0  & --1.120$\pm$0.071 &0.035 & 10.3&1.139$\pm$0.058&Y\\
m2.2 & 140.1812 & -77.3273 & 18.655 & 0.920 & --0.367$\pm$0.050 & -0.040& 8.6  & --1.105$\pm$0.067 &0.042 & 13.5&0.995$\pm$0.057&Y\\
m2.3 & 140.1606 & -77.3127 & 17.416 & 1.089 & --0.188$\pm$0.031 & 0.042& 18.6 & --1.226$\pm$0.035 &-0.072 & 31.0&1.041$\pm$0.029&Y\\
m2.4 & 140.1906 & -77.3123 & 15.972 & 1.152 & --0.095$\pm$0.017 & 0.022& 43.8 & --1.186$\pm$0.019 &-0.023 & 68.6&1.188$\pm$0.016&Y\\
m2.5 & 140.2596 & -77.3148 & 17.574 & 1.018 & --0.229$\pm$0.057 & 0.014& 8.2  & --1.181$\pm$0.076 &-0.028 & 11.8&1.003$\pm$0.062&Y\\
m2.6 & 140.2615 & -77.3141 & 19.335 & 0.813 & --0.286$\pm$0.055 & 0.095& 6.2  & --1.089$\pm$0.083 &0.054 & 8.1&0.912$\pm$0.072&Y\\
m2.7 & 140.1342 & -77.2990 & 15.474 & 1.169 & --0.012$\pm$0.014 & 0.066& 51.0 & --1.142$\pm$0.016 &0.024 & 81.7&1.146$\pm$0.013&Y\\
m2.8 & 140.1846 & -77.2997 & 18.650 & 0.914 & --0.401$\pm$0.039 & -0.074& 11.5 & --1.173$\pm$0.054 &-0.026 & 18.1&0.940$\pm$0.046&Y\\
m2.9 & 140.3156 & -77.3011 & 16.194 & 1.126 & --0.200$\pm$0.037 & -0.065& 20.3 & --1.138$\pm$0.044 &0.023 & 32.6&1.108$\pm$0.036&Y\\
m2.10 & 140.2847 & -77.2912 & 18.306 & 0.980 & --0.349$\pm$0.056 & -0.049& 6.0 & --1.188$\pm$0.078 &-0.039 & 10.7&1.042$\pm$0.065&Y\\
m2.11\tablenotemark{c} & 140.2880 & -77.2892 & 17.354 & 1.047 & --0.339$\pm$0.043 & -0.114&12.5 & --1.131$\pm$0.051 &0.024&19.8&1.125$\pm$0.043& Y\\
m2.12 & 140.3345 & -77.2901 & 16.721 & 1.089 & --0.074$\pm$0.058 & 0.102&9.0   & --1.190$\pm$0.070 &-0.032&13.4&1.056$\pm$0.056& Y\\
m2.13 & 140.2395 & -77.2789 & 17.038 & 1.004 & --0.271$\pm$0.021 & -0.070&29.5  & --1.153$\pm$0.027 &0.003&43.5&1.025$\pm$0.023& N\\
m2.14 & 140.2739 & -77.2794 & 17.812 & 1.030 & --0.337$\pm$0.033 & -0.076&16.0  & --1.185$\pm$0.043 &--0.033&24.2&1.039$\pm$0.036& Y\\
m2.15 & 140.2922 & -77.2691 & 17.651 & 1.039 & --0.262$\pm$0.034  & -0.013&15.3 & --1.173$\pm$0.043 &--0.020&24.5&1.066$\pm$0.036& Y\\
m2.16 & 140.3223 & -77.2586 & 18.033 & 1.031 & --0.071$\pm$0.047 & 0.208&10.7 & --1.091$\pm$0.051 &0.059&18.2&1.095$\pm$0.043& Y
\enddata
\tablenotetext{a}{S/N was measured in the interval 3894--3910\,\AA}
\tablenotetext{b}{S/N was measured in the interval 4240--4280\,\AA}
\tablenotetext{c}{This star was observed in both masks}
\end{deluxetable}
\clearpage

\end{document}